\documentclass[journal=jctcce,manuscript=article,layout=twocolumn]{achemso}
\setkeys{acs}{articletitle = true} 
\usepackage{graphicx}
\usepackage{subcaption}
\usepackage{color}
\usepackage{mathtools}
\usepackage{amsmath}
\usepackage[version=3]{mhchem} 

\usepackage{multirow}

\usepackage{siunitx}

\usepackage{times,mathptmx}
\usepackage{bm}
\usepackage{amsmath}
\usepackage{amsfonts}
\usepackage{times}  
\usepackage{txfonts} 
\usepackage{amssymb}
\usepackage{amsbsy}

\newlength{\starwidth}

\newcommand{\ket}[1]{| #1 \rangle}
\newcommand{\bra}[1]{\langle #1 |}
\newcommand{\braket}[2]{\langle #1 | #2 \rangle}

\newcommand{\etal}{{\it{et al.}}}
\newcommand{\brr}{\boldsymbol{r}}
\newcommand{\bR}{\boldsymbol{R}}

\newcommand{\bqq}{\boldsymbol{q}}

\newcommand{\diff}{ \! \mathrm{d}}

\newcommand{\abin}{\textit{ab initio }}

\definecolor{uoe-2}{RGB}{193, 0, 67}

\title{Excited Electronic States in Total Isotropic Scattering from Molecules}

\author{Nikola Zotev}
\affiliation{Centre for Science at Extreme Conditions and EaStCHEM School of Chemistry, University of Edinburgh, David
  Brewster Road, EH9 3FJ Edinburgh, UK}

\author{Andr\'es Moreno Carrascosa}
\affiliation{Centre for Science at Extreme Conditions and EaStCHEM School of Chemistry, University of Edinburgh, David
  Brewster Road, EH9 3FJ Edinburgh, UK}
  
\author{Mats Simmermacher}
\affiliation{Centre for Science at Extreme Conditions and EaStCHEM School of Chemistry, University of Edinburgh, David
  Brewster Road, EH9 3FJ Edinburgh, UK}
  
\author{Adam Kirrander}
\email{Adam.Kirrander@ed.ac.uk}
\affiliation{Centre for Science at Extreme Conditions and EaStCHEM School of Chemistry, University of Edinburgh, David
  Brewster Road, EH9 3FJ Edinburgh, UK}

\begin{document}


\maketitle

\begin{abstract}
Ultrafast x-ray scattering experiments are routinely analyzed in terms of the isotropic scattering component. Here we present an analytical method for calculating total isotropic scattering directly from \abin  two-electron densities of ground and excited electronic states. The method is generalized to compute isotropic elastic, inelastic, and coherent mixed scattering. The computational results focus on the potential for differentiating between electronic states and on the composition of the total scattering in terms of elastic and inelastic scattering. By studying the umbrella motion in the first excited state of ammonia, we show that the associated electron density redistribution leaves a comparably constant fingerprint in the total signal that is similar in magnitude to the contribution from the changes in molecular geometry.
\end{abstract}

\section{Introduction}
The X-ray Free-Electron Laser (XFEL) facilities that have emerged around the world in the last decade provide coherent and ultrashort x-ray pulses, whose peak brightness is more than ten orders of magnitude larger than for synchrotron radiation. With a pulse duration that rivals that of optical lasers, XFELs have greatly enriched the palette of experimental techniques used to study the most fundamental aspects of chemistry -- how molecules move,~\cite{Minitti2015} how chemical bonds are made or broken,~\cite{Ihee2015} and how electrons rearrange after interaction with light.~\cite{Lemke2013} One such powerful technique is non-resonant ultrafast x-ray scattering from gas-phase samples \cite{MinittiFD171,Minitti2015,BudarzJPB2015,Stankus2016,BucksbaumPRL2016,Yong2018,Ruddock2019}. In a pump-probe fashion, an ensemble of molecules is pumped by an optical laser to an excited electronic state, and the resulting photodynamics is probed via hard x-rays with varying delay time. Due to the fast nature of the dynamics and the small number of scattering molecules, these experiments require ultrashort pulse duration and large photon numbers currently only provided by XFELs.

It is not always recognized that gas-phase x-ray scattering has fundamental differences from x-ray crystallography. On account of the large average separation between molecules and the absence of a regular lattice, the gas-phase scattering is free from intermolecular interferences for all but the smallest values of the scattering vector.~\cite{Debye1927} The signal should thus be understood as an incoherent sum of scattering intensities from isolated molecules. It is on this single-molecule scale that quantum effects are most easily observed.  In contrast, the periodicity of crystals means that the signal at the Bragg peaks is strongly dominated by elastic scattering,~\cite{James} which, unlike the total scattering in gas phase, is a one-electron property.~\cite{BartellGavin1964,BartellGavin1965}
It is furthermore worth noting that x-ray crystallography is traditionally concerned with molecules in their thermal ground states, while the laser-induced dynamics in pump-probe experiments evolves on multiple electronic states, each characterized by its own distinctive electron distribution. It follows that the theoretical tools developed for and successfully applied in x-ray crystallography for more than a century are not always best suited for ultrafast x-ray scattering \cite{Techert2006,Techert2010,Techert2011,Northey2014,Northey2016,KirranderJCTC2016,KirranderWeber2017,Moreno2017,Moreno2017inelastic,Carrascosa2019total,Parrish2019}. 

Indeed, gas-phase scattering from ground state molecules initially, and, more recently, ultrafast x-ray scattering, have prompted the development of a number of algorithms that aim the accurate prediction of x-ray scattering starting directly from the \abin electronic structure of molecules. As gas-phase samples in thermal equilibrium are isotropic, a central question in these methodologies is that of rotational averaging. Wang and Smith have first suggested a direct analytical method for evaluating isotropic scattering intensities.~\cite{Wang1994} Various permutations of this approach have been devised since with the ultimate goal to reduce computational effort.~\cite{sarasola_closedform_1998, Thakkar2001,Crittenden2009} Alternative strategies that relay on numerical rotational averaging~\cite{Carrascosa2019total, Moreno2017, Hoffmeyer1998} or grid-density methods~\cite{Parrish2019} have also been suggested. Although it might not be immediately obvious, isotropic scattering is directly applicable to gas-phase ultrafast x-ray scattering despite the anisotropy imposed on the sample by the pump laser's polarisation. Separation of the isotropic from the anisotropic part of the experimental scattering signal can be achieved by means of a Legendre decomposition~\cite{Shipsey1992i,BenNun1997,Cao1998}, with the isotropic component equivalent to the outcome if the ensemble was fully isotropic~\cite{Lorenz2010,Baskin2006}. 

In this context, one of the challenges that ultrafast x-ray scattering is facing is how to accurately predict scattering intensities from the huge conformational space explored by the molecule during photochemical dynamics. This issue does not only hold the key to accurate simulations of ultrafast scattering but also couples to the problem of inversion of experimental data to molecular geometries via iterative procedures as done in x-ray crystallography. Currently available algorithms are often computationally too demanding for such a high-throughput task. In this article we extend the method developed by Crittenden and Bernard~\cite{Crittenden2009} for calculating isotropic scattering as a sum of spherical Bessel functions. We demonstrate the existence of a recursive relationship between the expansion coefficients that allows for a significant speed-up of the calculation, in addition to generalising the approach to an arbitrary angular momentum. We illustrate how the method scales with the level of theory and the basis set used. We show that it is applicable not only to elastic and total scattering but also to inelastic and coherent mixed terms between different electronic states.~\cite{Moreno2017inelastic,SimmermacherPRL2019}

We make use of our methodology to investigate the different components comprising the total scattering by a model multielectron two-state system, namely the ground and first excited states of ammonia along the umbrella normal mode. We demonstrate that in this particular case, the change in the scattering signal upon optical excitation cannot be simply attributed to changes in the molecular conformation only, and is in fact strongly influenced by the electron density redistribution in the excited state due to excitation.


\section{Theory}








For ultrafast x-ray scattering from photoexcited molecules, the differential scattering cross section per solid angle $\Omega$) can be shown to take the following form,~\cite{Moller2008}
\begin{align}\label{eq:start}
\begin{split}
\frac{\diff\sigma}{\diff\Omega}
& =
\left(\frac{\diff\sigma}{\diff\Omega}\right)_{\mathrm{Th}} 
W(\Delta\omega)
\\
& \times \int \diff t \ I (t) 
\bra{\Psi (t)}\sum_{i,j=1}^{N_{\mathrm{e}}} e^{\iota \bqq \cdot(\brr_i-\brr_j)}\ket{\Psi (t)}
\text{,}
\end{split}
\end{align}
where  $\Psi(t)$ is the time-dependent wavefunction of the molecule, $\bqq$ is the momentum transfer vector, and $r_i$ and $r_j$ are the position vectors of electrons $i$ and $j$, respectively. The detected signal is proportional to $(\diff\sigma/\diff\Omega)_{\mathrm{Th}}$, which is the differential Thomson scattering cross-section for a free electron which includes the polarization factor $|\mathbf{e}_1\cdot\mathbf{e}_2|$ of the incoming and scattered x-rays),~\cite{Thomson1906,Lorenz2010} and $W(\Delta\omega)$, the window function with a detection window defined by $\Delta\omega$. The braket notation implies integration over both the electronic and nuclear coordinates. Eq.~(\ref{eq:start}) is valid in the limit of a large detection window, \textit{i.e.}\ in the absence of energy resolution on the detector, as discussed in Ref.\ \citenum{SimmermacherPRL2019}. Note that, in deriving this equation, the high photon energies of the x-rays compared to the energy spectrum of a typical molecule allow us to apply the Waller-Hartree approximation\cite{Waller1929} and disregard the comparatively small changes in the photon energy.

Upon photoexcitation the molecular wave function, $\ket{\Psi (t)}$, is described by a Born-Huang expansion in the basis of the $N$ electronic eigenstates $\psi_{I} (\bar{\brr};\bar{\bR})$ accessed during the dynamics. The electronic eigenstates are functions of the electronic coordinates $\bar{\brr}$ of the $N_{\mathrm{e}}$ electrons in the molecule and depend parametrically on the molecular frame nuclear coordinates $\bar{\bR}$,
\begin{align}\label{eq:BornHuang}
\ket{\Psi (t)} = \sum_{\substack{I}=1}^{N}\ \ket{\chi^{\mathrm{rv}}_{I} (t)}\ \ket{\psi_{I}^{\vspace*{\starwidth}} (\bar{\bR})}\text{.}
\end{align}

In this expansion, each electronic state $I$ is multiplied by the corresponding time-dependent rovibrational nuclear wave packet $\ket{\chi^{\mathrm{rv}}_{I} (t)}$, which depends on the internal nuclear coordinates $\bar{\bR}$ and on the three Euler angles $\alpha\beta\gamma$, which relate the molecular and laboratory frames.
Using Eq.~(\ref{eq:BornHuang}), the differential scattering cross section becomes,
\begin{align}\label{eq:DSCS_long}
\begin{split}
\frac{\diff\sigma}{\diff\Omega}
& =
\left(\frac{\diff\sigma}{\diff\Omega}\right)_{\mathrm{Th}} 
W(\Delta\omega)
\\
& \times \sum_{I,J=1}^N
\int \diff t \ I (t)\ 
\bra{\chi^{\mathrm{rv}}_{I} (t)}
I_{IJ} (\bqq, \bar{\bR})
\ket{\chi^{\mathrm{rv}}_{J} (t)}
\text{,}
\end{split}
\end{align}

The key quantity in Eq.\ (\ref{eq:DSCS_long}) is $I_{IJ} (\bqq, \bar{\bR})$, which is the two-electron scattering matrix element given by,
\begin{align}\label{eq:Itotal}
I_{IJ} (\bqq, \bar{\bR}) &= \sum_{i,j}^{N_{\mathrm{e}}}\ \bra{\psi_{I}^{\vspace*{\starwidth}} (\bar{\bR)}} e^{\iota \bqq \cdot(\brr_{j} - \brr_{i})} \ket{\psi_{J}^{\vspace*{\starwidth}} (\bar{\bR})}\text{.}
\end{align}
Since terms with $i = j$ in Eq.\ (\ref{eq:Itotal}) reduce to the Kronecker delta $\delta_{IJ}$, the two-electron scattering matrix element can be written as,
\begin{align}\label{eq:Itotal_short}
I_{IJ}^{\vspace*{\starwidth}} (\bqq, \bar{\bR}) = N_{\mathrm{e}}\ \delta_{IJ} + I_{IJ}^{\prime} (\bqq, \bar{\bR})\text{,}    
\end{align}
where $I_{IJ}^{\prime} (\bqq, \bar{\bR})$ is the pure two-electron part of $I_{IJ}^{\vspace*{\starwidth}} (\bqq, \bar{\bR})$ with $i\neq j$. 
Going further, using the sifting property of the Dirac delta function, $\mathrm{exp}[\iota\bqq\cdot(\brr_i-\brr_j)]$ can be expressed in an integral form:
\begin{equation}
   e^{\iota\bqq\cdot(\brr_i-\brr_j)} =
   \iint d\brr_1\ d\brr_2\ e^{\iota\bqq\cdot(\brr_1-\brr_2)}
   \delta(\brr_1-\brr_i) \delta(\brr_2-\brr_j),
\end{equation}
so that the integral over the electronic coordinates in Eq.\ (\ref{eq:Itotal}) becomes,
\begin{align}\label{eq:I_elec}
\begin{split} 
    I_{IJ}(\bqq,\bar{\bR}) 
    & = N_\mathrm{e}\delta_{IJ} \\
    & +
    2\iint d\brr_1\ d\brr_2\  \rho_{IJ}^{(2)}(\brr_1, \brr_2, \bar{\bR})
    e^{\iota\bqq\cdot(\brr_1-\brr_2)},
\end{split}
\end{align}
where $\rho_{IJ}^{(2)}(\brr_1, \brr_2, \bar{\bR})$ is the expectation value of two-electron density operator $\hat{\rho} (\brr_{1}, \brr_{2}) = (1/2)\ \sum^{N_\mathrm{e}}_{\substack{i}} \sum^{N_\mathrm{e}}_{\substack{j\neq i}}\ \delta (\brr_{1} - \brr_i) \delta (\brr_{2} - \brr_j)$.~\cite{Helgaker} At that point, it is prudent to differentiate between the diagonal elements with respect to the electronic states, $I=J$, and off-diagonal (mixed) terms, with $I \neq J$. In the former case, $\rho_{II}^{(2)}(\brr_1, \brr_2, \bar{\bR})$ gives the probability of finding one of the electrons of the system in state $I$ at $\brr_1$, while another electron is at $\brr_2$. This term can be further separated into two contributions by expanding the two-electron density as a sum of products over one-electron density functions, $\rho_{IK}^{(1)}(\brr,\bar{\bR})$,
\begin{align} \label{eq:2RDF}
    \begin{split}
     N_\mathrm{e} + 2 \rho_{II}^{(2)}(\brr_1, \brr_2, \bar{\bR})
     &= 
     \rho_{II}^{(1)}(\brr_1,\bar{\bR}) \rho_{II}^{(1)}(\brr_2,\bar{\bR})
     \\
    &+ \sum_{K \neq I} \rho_{IK}^{(1)}(\brr_1,\bar{\bR}) \rho_{KI}^{(1)}(\brr_2,\bar{\bR}),
    \end{split}
\end{align}
which follows from insertion of the resolution of the identity in the basis of the electronic states.  The contribution to the total scattering from the first term in Eq.\ (\ref{eq:2RDF}) is the elastic scattering, while the contribution from the second term is the inelastic scattering. When $I \neq J$, the quantity $\rho_{IJ}^{(2)}(\brr_1, \brr_2, \bar{\bR})$ is referred to as the two-electron transition density function (\textit{i.e.}\ the diagonal part of the density matrix) in order to differentiate it from the case of $I=J$, which is simply known as the two-electron density function. The $I\neq J$ scattering terms play a critical role in coherent mixed scattering, which appears when there is a coherence between two electronic states \cite{Santra2012,Bennet2018,SimmermacherPRL2019}. This concludes the presentation of the fundamental theory.

\begin{figure}[htb!] 
\includegraphics[width=1\columnwidth]{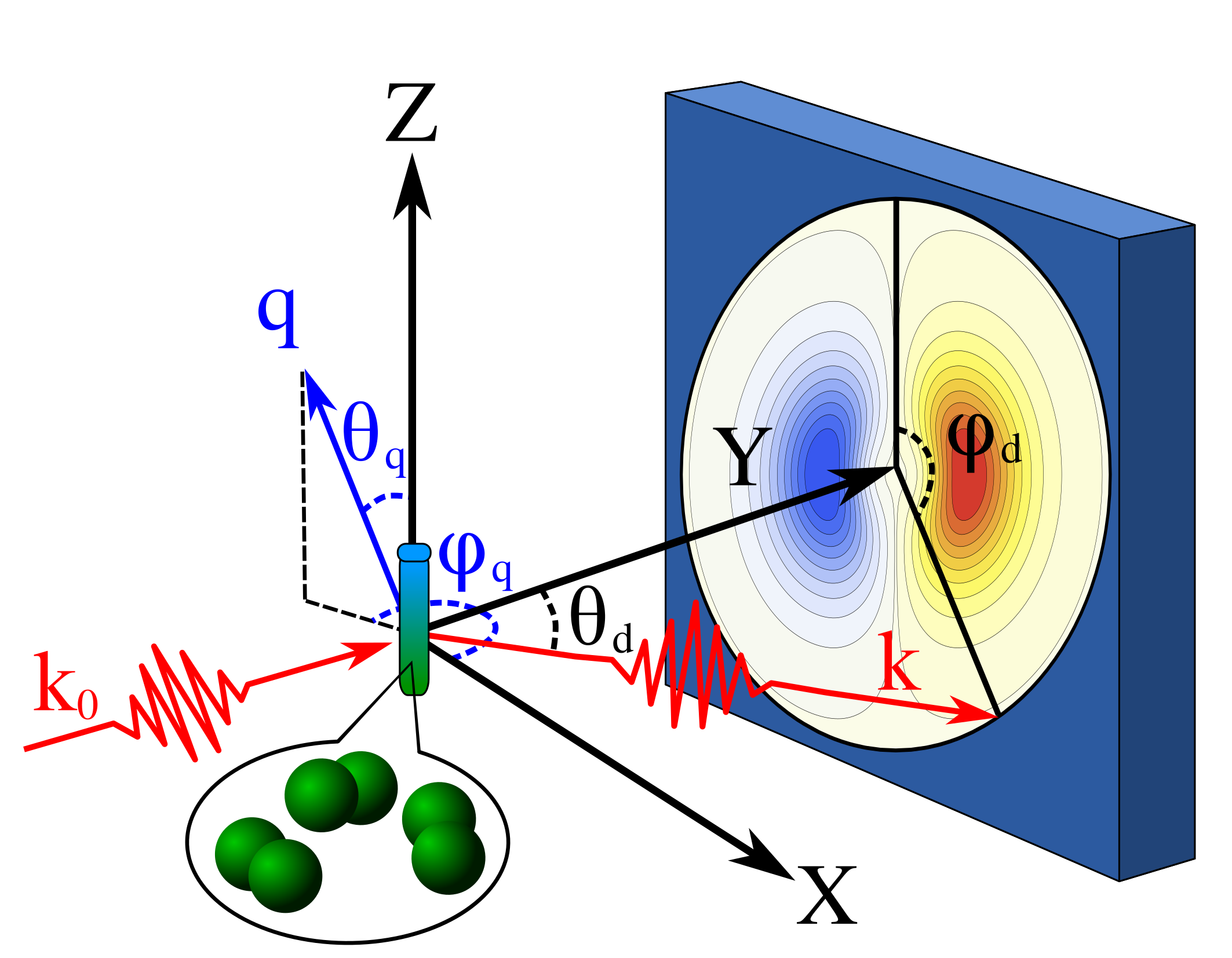} 
\caption{Illustration of the geometrical relations in X-ray scattering. The laboratory frame coordinate system XYZ is defined so that the incoming X-ray beam wave vector, $\boldsymbol{k}_0$, is aligned with the Y-axis, while the direction of the optical pump-laser polarisation points in the Z direction. After interaction with the sample, radiation is scattered in the direction  $\boldsymbol{k}$. The detector angles $\theta_d$ and $\phi_d$ are defined as the polar and the azimuthal angles of $\boldsymbol{k}$ with respect to the Y axis and the ZX plane, respectively. The momentum transfer vector, $\bqq=\boldsymbol{k}_0-\boldsymbol{k}$, forms the polar angle $\theta_q$ with the positive Z axis, and azimuthal the angle $\phi_q$ upon projection onto the XY plane. 
} \label{fig:polarCoord}
\end{figure}

The main focus of this article is an efficient methodology to calculate isotropic scattering signals. This is motivated by the central role that the isotropic scattering signal plays in the interpretation of experiments, irrespective of the degree of alignment in the sample.~\cite{Lorenz2010,Baskin2006} To appreciate this, we must consider the standard approach in gas-phase scattering experiments to decompose the observed signal in the basis of orthogonal Legendre polynomials, $P_\alpha(\cos \theta_q)$,~\cite{Shipsey1992i}
\begin{equation} \label{eq:legendre}
     \frac{\diff\sigma}{\diff\Omega}(\bqq)=\sum_{\alpha} P_{\alpha}( \cos \theta_q) S_{\alpha}(q),
\end{equation}
where  $\cos \theta_q$ is the component of the unit scattering vector in direction of the laser polarization axis shown in Fig.\ \ref{fig:polarCoord}. The angle $\theta_q$ is related to the detector angles by $\cos{\theta_q}=\sin{(\theta_d/2)} \cos{\delta} - \cos{(\theta_d/2)} \cos{\phi_d} \sin{\delta}$, where $\delta$ is angle between the directions of the laser polarisation and the X-ray beam propagation. In the case of a perpendicular pump-probe arrangement, \textit{i.e.} the case discussed here, when the laser polarisation axis is perpendicular to the X-ray propagation, this reduces to $\cos \theta_q = - \cos \frac{\theta_d}{2} \cos \phi_d$. It should be emphasised that the Legendre polynomials are functions of $\theta_q$ and not of the detector angles, $\theta_d$ and $\phi_d$.   The advantage of this approach is that the contribution from the internal and external molecular degrees of freedom can be separated out, as will be shown below. It is important to point out that in many pump-probe gas phase scattering experiments, the detected signal is rarely fully isotropic on account of the preferential excitation of the molecules whose transition dipole moments align with the polarization axis of the linearly polarized pump laser.~\cite{Yong2018} 


A remarkable property of the decomposition in Eq.~(\ref{eq:legendre}) is that the entire information for the internal degrees of freedom can be extracted from any of the components, $S_\alpha(q)$, or a combination of them.~\cite{Biasin2018} While in the high-order terms, the internal dynamics is mingled with information about the rotational wavepacket, the zeroth order term in the Legendre expansion can be analysed solely from the point of view of the internal molecular degrees of freedom. It is exactly equivalent to the scattering from a fully isotropic ensemble and consistent with the Debye scattering formalism (see Supplementary Information for derivation). Using the orthogonality of the Legendre polynomials and that $P_0(\cos\theta_q)=1$, the zeroth-order term, which is henceforth referred to as isotropic scattering, is given by,
\begin{equation} \label{eq:iso1}
\begin{split}
   S_0(q) &= \frac{1}{4\pi} \int_0^{2\pi} 
   \diff\phi_q \int_0^\pi 
   \diff\theta_q \sin\theta_q  
   \frac{\diff\sigma}{\diff\Omega}(\bqq)\\ 
   &=  \Big \langle  \frac{\diff\sigma}{\diff\Omega}(\bqq) \Big \rangle_{\theta_q \phi_q},
\end{split}
\end{equation}
where the average is taken over the angular coordinates of the momentum transfer vector. With reference to the full expression for the differential scattering cross-section in Eq.~(\ref{eq:DSCS_long}), only the two-electron scattering matrix element, $I_{IJ} (\bqq, \bar{\bR})$, depends on the scattering vector. Furthermore, the average renders the scattering signal independent of the orientation of the molecule in space, meaning that $I_{IJ}(q,\bar{\bR} )=\big \langle I_{IJ}(\bqq,\bar{\bR}) \big \rangle_{\theta_q \phi_q}$ does not depend on the Euler angles (detailed derivation included in Supplementary Information). That allows for a separate integration of the rotational wave packets, resulting in a simple scaling factor, which is equal to one for $I=J$,


\begin{align}\label{eq:iso2}
\begin{split}
S_0(q)
& =
\left(\frac{\diff\sigma}{\diff\Omega}\right)_{\mathrm{Th}}
W(\Delta\omega) \\
& \times
\sum_{I,J}
\int \diff t \ I (t) 
\braket{ \Theta_I^{\text{r}}}{\Theta_J^{\text{r}} }
\bra{\chi_I^{\mathrm{v}} }
    I_{IJ}(q,\bar{\bR} )
\ket{  \chi_J^{\mathrm{v}} }
\text{,}
\end{split}
\end{align}
where the vibrational, $\ket{\chi^\mathrm{v}}$, and rotational, $\ket{\Theta^\mathrm{r}}$, components of the rovibrational wave packet have been made explicit, and their dependence on time have been omitted for brevity. It follows that the isotropic component of the experimental scattering signal can be analyzed, as if the entire ensemble showed a fully isotropic rotational distribution.


\section{Methods} \label{sec:methods}

We now turn our attention to the analytical evaluation of the isotropic differential scattering cross section. We will seek a solution for the rotational-average two-electron scattering matrix elements, $I_{IJ}(q,\bar{\bR} )=\big \langle I_{IJ}(\bqq,\bar{\bR}) \big \rangle_{\theta_q \phi_q}$, which are a prerequisite for a more detailed description later on, that involves the role of nuclear motion. From Eq.~(\ref{eq:I_elec}), we will need to evaulate the expression:
\begin{align}\label{eq:I_elec2}
\begin{split} 
    I_{IJ}(q,\bar{\bR}) 
    & = N_\mathrm{e}\delta_{IJ} \\
    & +
    2 \Bigg \langle 
    \iint d\brr_1\ d\brr_2\  \rho_{IJ}^{(2)}(\brr_1, \brr_2, \bar{\bR})
    e^{\iota\bqq\cdot(\brr_1-\brr_2)}
    \Bigg \rangle
    ,
\end{split}
\end{align}
In the absence of energy resolution, \textit{i.e.}\ the standard set-up for current time-resolved scattering experiments, the separation of the total scattering into elastic and inelastic components is not needed, but it is instructive to show that our methodology is applicable to all four cases: total, elastic, inelastic and coherent mixed. Thus, we can split the expression for total scattering ($I=J$) using Eq.\ (\ref{eq:2RDF}):
\begin{align}\label{eq:I_elec3}
\begin{split} 
    I_{II} (q,\bar{\bR}) 
    &  = \sum_K^{\infty} 
    \Bigg \langle 
    \iint d\brr_1\ d\brr_2\ \\
    &
    \times  \rho_{IK}^{(1)}(\brr_1,\bar{\bR}) \rho_{KI}^{(1)}(\brr_2,\bar{\bR}) 
    e^{\iota\bqq\cdot(\brr_1-\brr_2)}
    \Bigg \rangle
    ,
\end{split}
\end{align}
where terms with $K=I$ are elastic, while $K \neq I$ are inelastic. The key quantities in  Eq.\ (\ref{eq:I_elec2}) and Eq.\ (\ref{eq:I_elec3}) are the one- and two-electron density functions, $\rho_{IJ}^{(1)}(\brr_1)$ and $\rho_{IJ}^{(2)}(\brr_1, \brr_2)$, which can be expressed as weighted products of molecular orbitals (dropping the parametric dependence on the nuclear coordinates),~\cite{Helgaker}
\begin{equation} \label{eq:1RDM}
     \rho_{IJ}^{(1)}(\brr_1) = \sum_{ab}^{N_{\mathrm{MO}}} D_{ab}^{IJ} \phi_a(\brr_1) \phi_b(\brr_1).
\end{equation}
and
\begin{equation} \label{eq:2RDM}
    \rho_{IJ}^{(2)}(\brr_1, \brr_2) = \frac{1}{2} \sum_{abcd}^{N_{\mathrm{MO}}}  d_{abcd}^{IJ} \phi_a(\brr_1) \phi_b(\brr_1)  \phi_c(\brr_2) \phi_d(\brr_2),
\end{equation}
where the indices run over all occupied molecular orbitals, $N_{\mathrm{MO}}$. The terms $D_{ij}^{II}$ and $d_{ijkr}^{II}$ are referred to as the elements of the one- and two-electron reduced density matrix (1- and 2-RDM), respectively. If $I \neq J$, they are known as one- and two-electron reduced transition density matrix elements.

Introducing Eq.\ (\ref{eq:1RDM}) and Eq.\ (\ref{eq:2RDM}) into the expression for the isotropic scattering, Eq.\ (\ref{eq:I_elec2}) and Eq.\ (\ref{eq:I_elec3}), shows that the efficient \abin solution to the isotropic scattering problem  requires the evaluation of integrals of the type,
\begin{align}
    \begin{split}
        K_{IJ}(q) &=
        \Bigg \langle
        \iint \diff \brr_1 \, \diff \brr_2  
        \sum_{abcd}^{N_{\mathrm{MO}}} z_{abcd}^{IJ} \\
        & \times
        \phi_a(\brr_1) \phi_b(\brr_1) \phi_c(\brr_2) \phi_d(\brr_2)
        e^{\iota\bqq\cdot(\brr_1-\brr_2)}
        \Bigg \rangle_{\theta_q \phi_q},
    \end{split}
\end{align}

where 

\begin{equation}
    z_{abcd}^{IJ} =    
    \begin{cases}
        d_{abcd}^{II} ,             & \text{total scattering } (I=J)\\
        D_{ab}^{II} D_{cd}^{II},    & \text{elastic scattering } (I=J) \\
        D_{ab}^{IJ} D_{cd}^{JI},    & \text{inelastic scattering } (I\neq J) \\
        d_{abcd}^{IJ}.              & \text{coherent mixed } (I\neq J) \\
    \end{cases}
\end{equation}

A common strategy in molecular electronic structure theory is to expand the orbitals as a weighted sum of $N_{\mathrm{bf}}$ primitive Cartesian Gaussian-type orbitals (GTOs),
\begin{align}
    \begin{split}
            &\phi_a(\brr_1) 
            = \sum_i^{N_{\mathrm{bf}}} M_i^{(a)} g_i(\brr) \\
            &= \sum_i^{N_{\mathrm{bf}}} M_i^{(a)} (x-A_{i,x})^{l_i} (y-A_{i,y})^{m_i} (z-A_{i,z})^{n_i} 
            e^{-\gamma_i(\brr-\boldsymbol{A_i})^2}
    \end{split}
\end{align}
where $\boldsymbol{A_i}$ is the centre of the $i^{th}$ primitive Gaussian and $M_i^{(a)}$ the molecular orbital expansion coefficient. In the case of contracted Gaussian functions, $M_i^{(a)}$ is premultiplied by a contraction coefficient. The sums of $l_i$, $m_i$ and $n_i$ specify the orbital angular momentum. The integral $K_{IJ}$ then takes the form,
\begin{align} \label{eq:Kintegral}
    \begin{split}
    K_{IJ}(q) = \sum_{ijkr}^{N_{\mathrm{bf}}} Z_{ijkr}^{IJ}
    \Big \langle
    J_{ij}(\bqq) J_{kr}^{*}(\bqq)
    \Big \rangle_{\theta_q \phi_q},
    \end{split}
\end{align}
with the two-electron charge density $Z_{ijkr}^{IJ}=\sum_{abcd}^{N_{\mathrm{MO}}} z_{abcd}^{IJ} M_i^{(a)} M_j^{(b)} M_k^{(c)} M_r^{(d)}$, and where we have labeled the Fourier integrals over $\brr_1$ and $\brr_2$ as $J_{ij}(\bqq)$ and $J_{kr}^{*}(\bqq)$, respectively,
\begin{align} \label{eq:Kintegral-J1}
    \begin{split}
    J_{ij}(\bqq) = \int \diff \brr_1 g_i(\brr_1) g_j(\brr_1) e^{\iota\bqq\cdot\brr_1}
    \end{split}
\end{align}
and 
\begin{align} \label{eq:Kintegral-J2}
    \begin{split}
     J_{kr}^{*}(\bqq) = \int \diff \brr_2 g_k(\brr_2) g_r(\brr_2)
    e^{-i\bqq\cdot\brr_2}.
    \end{split}
\end{align}



Eq.\ (\ref{eq:Kintegral}) reveals that the calculation of the \abin isotropic scattering requires the evaluation of $N_{\mathrm{bf}}^4$ angular integrals. This is a formidable computational challenge, even for the smallest molecules with an adequate basis set. However, schemes for the efficient evaluation of such integrals have been proposed before~\cite{Wang1994,Crittenden2009} and make use of the properties of the Gaussian functions and their analytical Fourier transforms. The first step is to express the angular momentum properties of the product of two primitive Gaussians, $g_i$ and $g_j$ in a derivative form,~\cite{McMurchie1978} 
\begin{align} \label{eq:McMurchie}
\begin{split}
    \Pi_{ij}(\brr) 
    & = g_i(\brr) g_j(\brr) \\
    & = E_{ij} \sum_{L_1=0}^{l_i+l_j}  \sum_{M_1=0}^{m_i+m_j}  \sum_{N_1=0}^{n_i+n_j}
    \Lambda_{L_1}^{l_i l_j}
    \Lambda_{M_1}^{m_i m_j}
    \Lambda_{N_1}^{n_i n_j} \\
    &\times
    \left(\frac{\partial}{\partial P_x} \right)^{L_1}
    \left(\frac{\partial}{\partial P_y} \right)^{M_1}
    \left(\frac{\partial}{\partial P_z} \right)^{N_1}
    e^{-\gamma_P(\brr-\boldsymbol{P})^2},
\end{split}
\end{align}

where we do not explicitly show the dependence of the McMurchie-Davidson coefficients, $\Lambda_{L}^{l_i l_j}(x_i,x_j,\gamma_i, \gamma_j)$, on the Cartesian components of the centres of the Gaussian functions and Gaussian exponents. The expression also exploits that the product of two Gaussians functions is a new Gaussian function, leading to the following definitions that appear in Eq.\ (\ref{eq:McMurchie}),
\begin{eqnarray}
    \gamma_P &=& \gamma_i+\gamma_j, \nonumber \\
    E_{ij} &=& e^{-(\gamma_i\gamma_j/\gamma_P)(\boldsymbol{A}-\boldsymbol{B})^2}, \\
    \boldsymbol{P} &=& (\gamma_i\boldsymbol{A}+\gamma_j\boldsymbol{B})/\gamma_P. \nonumber
\end{eqnarray}

The McMurchie-Davidson expansion is widely used in computational chemistry programme packages to express integrals of Gaussian functions with high angular momentum as derivatives of $s$-type Gaussian integrals. Their utility stems from the existence of a recursive relationship between the coefficients, which enables their rapid evaluation. Substituting $\Pi_{ij}(\brr)$ into the expression for $J_{ij}(\bqq)$ and taking the derivative in front of the Fourier integral, results in,
\begin{align}
    \begin{split}
        J_{ij} & (\bqq)
        =\int \diff\brr 
        \, \Pi_{ij}(\brr) e^{\iota\bqq\cdot\brr}
        \\ = &
        E_{ij} \sum_{L_1=0}^{l_i+l_j}  \sum_{M_1=0}^{m_i+m_j}  \sum_{N_1=0}^{n_i+n_j}
        \Lambda_{L_1}^{l_i l_j}
        \Lambda_{M_1}^{m_i m_j}
        \Lambda_{N_1}^{n_i n_j} \\
        &\times
        \left(\frac{\partial}{\partial P_x} \right)^{L_1}
        \left(\frac{\partial}{\partial P_y} \right)^{M_1}
        \left(\frac{\partial}{\partial P_z} \right)^{N_1}
        \int \diff\brr 
        e^{-\gamma_P(\brr-\boldsymbol{P})^2}
        e^{\iota\bqq\cdot\brr}.
    \end{split}
\end{align}

The Fourier Transform of the $s$-type Gaussian function can be evaluated using the Fourier shift property,
\begin{equation}
    \int \diff\brr 
        e^{-\gamma_P(\brr-\boldsymbol{P})^2}
        e^{\iota\bqq\cdot\brr}
        =
        \left( \frac{\pi}{\gamma_P} \right)^{3/2}
        e^{-q^2/4\gamma_P}
        e^{\iota\bqq\cdot\boldsymbol{P}}.
\end{equation}
Having analytically performed the Fourier Transform of the product of two arbitrarily GTOs, the full expression for $J_{ij}(\bqq) J_{kr}^{*}(\bqq)$ prior to integrating out the angular dependence becomes,
\begin{align} \label{eq:big}
    \begin{split}
        &
        J_{ij}(\bqq) J_{kr}^{*}(\bqq)
        =
        \int \diff\brr_1 
        \Pi_{ij}(\brr_1) e^{\iota\bqq\cdot\brr_1}
        \int \diff\brr_2
        \Pi_{ij}(\brr_2) e^{-i\bqq\cdot\brr_2}
        \\
        &=
        \frac{\pi^3 E_{ij} E_{kr}}{(\gamma_P \gamma_Q)^{3/2}}
        e^{-q^2(1/\gamma_Q+1/\gamma_P)/4}
        \\ 
        & \times
        \sum_{L_1=0}^{l_i+l_j}  \sum_{M_1=0}^{m_i+m_j}  \sum_{N_1=0}^{n_i+n_j}
        \sum_{L_2=0}^{l_k+l_r}  \sum_{M_2=0}^{m_k+m_r}  \sum_{N_2=0}^{n_k+n_r} 
        \\ & \times
        \Lambda_{L_1}^{l_i l_j}
        \Lambda_{M_1}^{m_i m_j}
        \Lambda_{N_1}^{n_i n_j}
        \Lambda_{L_2}^{l_k l_r}
        \Lambda_{M_2}^{m_k m_r}
        \Lambda_{N_2}^{n_k n_r}
        F_{L_1 M_1 N_1}^{L_2 M_2 N_2}(\bqq,\boldsymbol{P},\boldsymbol{Q}),
    \end{split}
\end{align}
where
\begin{align}
    \begin{split} \label{eq:Fintegral}
        F_{L_1 M_1 N_1}^{L_2 M_2 N_2}
        &
        (\bqq,\boldsymbol{P},\boldsymbol{Q})
        =
        \left(\frac{\partial}{\partial P_x} \right)^{L_1} \!
        \left(\frac{\partial}{\partial P_y} \right)^{M_1} \!
        \left(\frac{\partial}{\partial P_z} \right)^{N_1} \! \\
        & \times
        \left(\frac{\partial}{\partial Q_x} \right)^{L_2} \!
        \left(\frac{\partial}{\partial Q_y} \right)^{M_2} \!
        \left(\frac{\partial}{\partial Q_z} \right)^{N_2} \!
        e^{\iota\bqq\cdot(\boldsymbol{P}-\boldsymbol{Q})}.
    \end{split}
\end{align}

Introducing $\boldsymbol{H}=\boldsymbol{P}-\boldsymbol{Q}$ and the combined angular momentum quantum numbers, $L=L_1+L_2$, $M=M_1+M_2$ and $N=N_1+N_2$, Eq.\ (\ref{eq:Fintegral}) can be written as,
\begin{align} \label{eq:Fintegral2}
    \begin{split}
        F_{L_1 M_1 N_1}^{L_2 M_2 N_2} &
        (\bqq,\boldsymbol{P},\boldsymbol{Q})
        = (-1)^{L_2+M_2+N_2}
        \\
        & \times
        \left(\frac{\partial}{\partial H_x} \right)^L \!
        \left(\frac{\partial}{\partial H_y} \right)^M \!
        \left(\frac{\partial}{\partial H_z} \right)^N \!  
        e^{\iota\bqq\cdot\boldsymbol{H}} \\
        & = (-1)^{L_2+M_2+N_2} F_{L M N}(\bqq,\boldsymbol{H}).
    \end{split}
\end{align}

It should be recognized that, if there was no need to perform the rotational average calculation, \textit{i.e.}\ for scattering in the molecular frame, the derivatives in Eq.\ (\ref{eq:Fintegral2}) trivially evaluate to $(iq_x)^L (iq_y)^M (iq_z)^N \mathrm{exp}[i\bqq\cdot\boldsymbol{H}]$. However, even in the case of isotropic scattering the quantity $\big \langle J_{ij}(\bqq) J_{kr}^{*}(\bqq) \big \rangle_{\theta_q \phi_q}$ expressed in its current form has a relatively simple analytic solution. Resolving the angular integrals, which only affects $e^{\iota\bqq\cdot\boldsymbol{H}}$ results in,
\begin{align}
    \begin{split} \label{eq:sinc}
    \Big \langle
    F_{L M N} & (\bqq,\boldsymbol{H}) 
    \Big \rangle_{\theta_q \phi_q} \\
    &=
    \left(\frac{\partial}{\partial H_x} \right)^{L}
    \left(\frac{\partial}{\partial H_y} \right)^{M}
    \left(\frac{\partial}{\partial H_z} \right)^{N}   
    \frac{\sin qH}{qH}.
    \end{split}
\end{align}

Solutions to the the equation above are discussed by Wang \etal~\cite{Wang1994}, where it is given as a four dimensional sum over trigonometric functions scaled by precalculated numerical factors. Here, we follow more closely the approach suggested by Crittenden \etal~\cite{Crittenden2009}, who calculated the result analytically for a limited number of angular momenta as a sum of spherical Bessel functions. In contrast to their approach, we recognize the existence of a recursive relationship between the expansion coefficients, which allows for a fast calculation and handling of arbitrarily large angular momenta. In the simple case when $H<\epsilon_{\text{cut}}$, the exponential in Eq.\ (\ref{eq:Fintegral2}) is approximately unity and Eq.\ (\ref{eq:sinc}) takes the form,
\begin{align}
    \begin{split}
    \Big \langle
    F_{L M N}(\bqq,\boldsymbol{H}) 
    \Big \rangle_{\theta_q \phi_q} 
    & =
    i^{L+M+N} 
    \Big \langle
    q_x^L q_y^M q_z^N
    \Big \rangle_{\theta_q \phi_q} \\
    & = B_{LMN}
    (iq)^{L+M+N},
    \end{split}
\end{align}
where 
\begin{equation}
B_{LMN}  = \Big \langle\sin^{L+M}\theta_q \cos^N \theta_q \cos^L \phi_q \sin^M \phi_q \Big \rangle_{\theta_q \phi_q}
\end{equation}
is a numerical constant. As discussed above, in the case when $H \geq \epsilon_{\text{cut}}$, the evaluation of the derivatives relies on the properties of the spherical Bessel functions, $j_\beta (qH)$,
\begin{align} \label{eq:bessel}
    \begin{split}
        \Big \langle
        & F_{L M N} (\bqq,\boldsymbol{H}) 
        \Big \rangle_{\theta_q \phi_q} \\
        & =
        \left(\frac{\partial}{\partial H_x} \right)^{L}
        \left(\frac{\partial}{\partial H_y} \right)^{M}
        \left(\frac{\partial}{\partial H_z} \right)^{N}   
        \frac{\sin qH}{qH}
        \\ &=
        \sum_{p=0}^{L}
        \sum_{s=0}^{M}
        \sum_{t=0}^{N}
        a_{L}^{p}(H_x) b_{M}^{s}(H_y) c_{N}^{t}(H_z) \left( \frac{q}{H} \right)^\beta j_\beta (qH),
    \end{split}
\end{align}
where $\beta=\text{min}[(L+M+N-p-s-t)/2]+p+s+t$, with $\text{min}[]$ denoting the least integer greater than or equal to the quantity in the brackets. The coefficients $a_{L}^{p}$, $b_{M}^{s}$ and $c_{N}^{t}$ are related to the Hermite polynomials and obey the following recursive relation (here given for $a_{L}^{p}$),
\begin{equation}
    a_{L}^{p}(H_x)=
    \begin{cases}
        1,                                  & L=0,\ p=0 \\
        0,                                  & L=1,\ p=0 \\
        -H_x,                               & L=1,\ p=1 \\
        -a_{L-2}^{0}(L-1),                    & L>1,\ p=0 \\
        -a_{L-1}^{p-1}H_x-a_{L-2}^{p}(L-1).    & L>1,\ p>0 \\
    \end{cases}
\end{equation}

The implementation of a sensible cut-off value, $\epsilon_{\text{cut}}$, is essential for the numerical stability of the algorithm, which might otherwise be affected by prohibitively large values of the $(q/H)^{\beta}$ factor in Eq.\ (\ref{eq:bessel}). The efficient procedure for evaluating Eq.\ (\ref{eq:sinc}) makes use of the recursive relationship for the coefficients $a_{L}^{p}$, $b_{M}^{s}$ and $c_{N}^{t}$ as well as of the recursive formula for the spherical Bessel functions. In addition to that, it is of paramount importance for the computational efficiency to take into account the symmetries $\big \langle J_{ij}(\bqq) J_{kr}^{*}(\bqq) \big \rangle_{\theta_q \phi_q} = \big \langle J_{kr}(\bqq) J_{ij}^{*}(\bqq) \big \rangle_{\theta_q \phi_q}$ and $J_{ij}(\bqq)=J_{ji}(\bqq)$, which together result in a speed-up factor of approximately 8. Another important simplification stems from careful consideration of the contraction scheme of the basis set used. If a given primitive GTO is a part of multiple contractions, the corresponding integral should be performed only once and the constant $Z_{ijkr}^{IJ}$ needs to be modified to reflect the combined contribution of this primitive to the molecular orbitals. The final trick for improving the computational performance is to treat together all GTOs whose centres and exponents are the same. Careful examination of these cases shows that they ultimately lead to the same values of $\boldsymbol{H}$ and can only differ in their angular momentum numbers $l$, $m$ and $n$. Treating them together results in a family of $\big \langle F_{L M N}(\bqq,\boldsymbol{H}) \big \rangle_{\theta_q \phi_q} $ integrals, for which  most of the terms in Eq.\ (\ref{eq:bessel}) are shared. In addition, a global cut-off linked to the relative size of $z_{abcd}^{IJ}$ could significantly speed up the calculation, at the expense of an effective decrease of the total number of electrons integrated. We have noticed that allowing for 0.1\% electron density loss can lead to an approximate speed-up factor of two in most molecules explored without significant effect on the results.

\section{Results and Discussion}

\subsection{Benchmarking and scaling}

Our method is extensively tested and validated in a series of \abin calculations for the ammonia molecule, \ce{NH3}, varying the basis set and size of the active space. All \abin calculations are performed with the MOLPRO electronic structure software package~\cite{MOLPRO-WIREs, MOLPRO}. We chose ammonia because it has previously been used by Hoffmeyer \etal\ \cite{Hoffmeyer1998} to illustrate the importance of multiconfigurational wavefunctions in total x-ray and electron scattering. Although, as shown above, our methodology also encompasses individual inelastic and coherent mixed transitions, we focus on total and elastic scattering here. The discussion of inelastic scattering is limited to the cumulative sum of all inelastic transitions, which is given by the difference between total and elastic scattering. The richness of the information encoded in the inelastic and coherent mixed terms will be the target of a follow-up publication.

\begin{figure*}     
    \centering
    \begin{subfigure}{1\columnwidth} \label{fig:convergence_total}
        \includegraphics[width=\columnwidth]{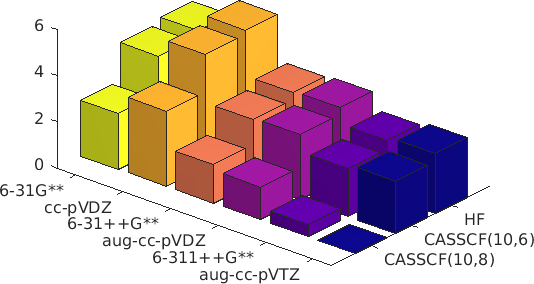}
        \caption{\ce{NH3} total scattering }%
    \end{subfigure}\hfill%
    \begin{subfigure}{1\columnwidth} \label{fig:convergence_elastic}
        \includegraphics[width=\columnwidth]{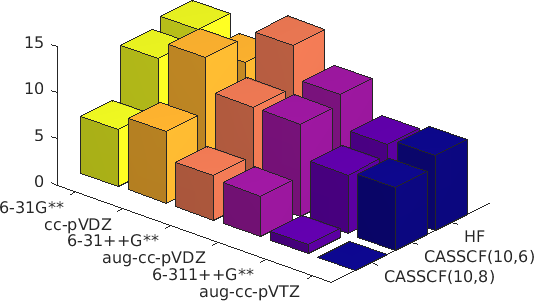}
        \caption{\ce{NH3} elastic scattering }%
    \end{subfigure}\hfill%
    \caption{Convergence of the total and elastic X-ray scattering signals of \ce{NH_3} computed with HF, CASSCF(10,6) and CASSCF(10,8) using various basis sets. The height of the bars represent the integral of the percentage intensity change with respect to CASSCF(10,8)/aug-cc-pVTZ total and elastic scattering for the range $0 \leq q \leq 11.34$~\AA$^{-1}$.}
    \label{fig:convergence}
\end{figure*}

Fig.~\ref{fig:convergence} compares total and elastic scattering at different levels of theory, namely Hartree-Fock (HF) and Complete Active Space Self-Consistent Field (CASSCF) with 6 and 8 active orbitals and all electrons active, \textit{i.e.}\ CASSCF(10,6) and CASSCF(10,8), respectively. The basis sets include Pople's and Dunning's correlation-consistent basis sets with double-zeta, double-zeta plus diffuse functions and triple-zeta plus diffuse functions. Calculations with STO-3G minimal basis set are also performed and are included in SI Table 1 and SI Table 2, which summarise the benchmarking results. The geometry used in all calculations is optimized at the CASSCF(10,8)/aug-cc-pVTZ level of theory. The results are first calculated as the fractional signal change (commonly expressed in percent),
\begin{equation} \label{eq:deltaS}
    \Delta S(q) = \frac{I(q)-I_{\mathrm{ref}}(q)}{I_{\mathrm{ref}}(q)},
\end{equation}
where the reference, $I_{\mathrm{ref}}$, is either the total or the elastic scattering computed at the CASSCF(10,8)/aug-cc-pVTZ level. The results presented in Fig.~\ref{fig:convergence} are given as the integral of the absolute values in the range $[q_{min},\ q_{max}] = [0,\ 11.34]$~\AA$^{-1}$,
\begin{equation} \label{eq:E}
    E=\int_{q_{min}}^{q_{max}} \diff q\ \big| \% \Delta S(q) \big|.
\end{equation}
We choose to use the fractional signal change (in percentage), $\% \Delta S(q)$,  to ensure that the difference at each momentum transfer vector is relative to the absolute value of the intensity at this point. It is worth noting that the signals in the limit of large $q$ for the total and elastic scattering differ by the number of the electrons in the molecule, which results in a larger value in the denominator in Eq.\ (\ref{eq:deltaS}) and a smaller integral in Eq.\ (\ref{eq:E}) for total scattering. 

It is clear that the two families of basis functions are comparable at a given level of theory. Interestingly, the Pople's basis sets with the smaller active space and Hartree-Fock seem to be closer to the reference CASSCF(10,8)/aug-cc-pVTZ compared to the correlation-consistent basis sets. It is possible that the split-valence basis sets provide a better description in cases where the active space is not sufficient to capture the static electron correlation adequately. In addition, we found that STO-3G is largely unsuitable for scattering calculations (SI Table 1 and SI Table 2). In fact, it is comparable to the Independent Atom Model (IAM),~\cite{KirranderJCTC2016} which gives $E=17.2$~\AA$^{-1}$ and $E=32.8$~\AA$^{-1}$ for total and elastic scattering. Overall, in both Pople's and Dunning's basis sets, the convergence is smooth with increasing the number of basis functions. The most significant effect is seen from going from the smaller  CASSCF(10,6) to the larger CASSCF(10,8) active space, which can be attributed to the effects of electron correlation on the electron density of the system. We note that the total scattering is two-electron property and will be significantly affected by electron correlation. However, it is clear that elastic scattering convergence is equally affected suggesting that the electron density relaxation associated with static correlation, implicit in the CASSCF calculations, is a major factor in that case. For accurate total scattering calculation, \abin methods that capture dynamic correlation are desirable.

\begin{figure} 
  \centering
    \includegraphics[width=1\linewidth]{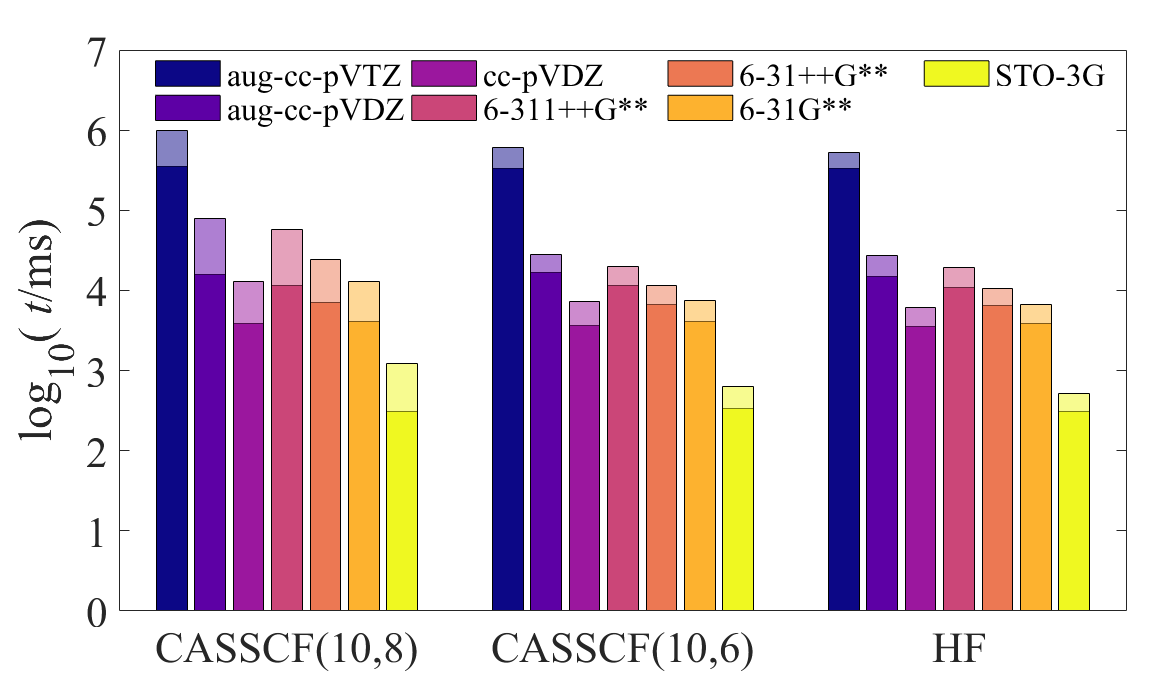}
    \caption{Comparison of the computational time required for the calculation of the isotropic elastic and total x-ray scattering signals in \ce{NH3}. Different \abin methods are considered as well as basis sets. Note that the computational scaling is expressed as a logarithm of the CPU time in milliseconds. The solid part of the bars represent the elastic scattering, whereas the shaded area at the top shows the extra time required to compute the total scattering with the same level of theory and basis set.}
    \label{fig:time}
 \end{figure}

Fig.~\ref{fig:time} compares the computational time required for the calculations discussed above. Generally, the calculations scale with the fourth power of the number of basis functions, $N_{\mathrm{bf}}^4$, as all distinctive Fourier transforms over four basis functions need to be considered if they are not related by permutational symmetry. As discussed above, grouping together integrals with equivalent origin in Cartesian space and exponents partially offsets this scaling. Practically, as the number of atoms increases, so does the number of unique centres so that the speed up for large molecules could be small. In the case of elastic scattering, the two-electron charge density $Z_{ijkr}^{IJ}$ can be expressed as two independent pairs $Z_{ij}^{IJ}$ and $Z_{kr}^{IJ}$. After precomputing these results, they can easily be accessed in the calculation with no added computational cost. Hence, the elastic scattering is largely independent of the level of theory or active space used. For the total scattering, the two-electron charge density remains a function of all four GTOs, so that its {\it{on-the-fly}} calculation increases the computational time compared to the elastic scattering in a manner that scales with the number of active orbitals. The scalings for specific inelastic and coherent mixed terms are equivalent to elastic and total scattering, respectively. Overall, we note that the computational time required per scattering calculation with the largest basis set used in this work is three orders of magnitude more expensive than a simple calculation with a minimal basis set. This scaling is of great practical importance for choosing an optimal method when considering scattering from a wide range of molecular geometries along a reaction coordinate or for the purpose of iterative inversion of experimental data. Given the comparatively small overhead cost of total scattering, it is sensible to adopt this type of calculation \textit{in lieu} of the common approach of calculating total scattering as a sum of elastic scattering and tabulated inelastic corrections.~\cite{Ruddock2019}

\begin{figure*}[htb]
    \begin{subfigure}{1\columnwidth}
        \centering
        \includegraphics[width=1\linewidth]{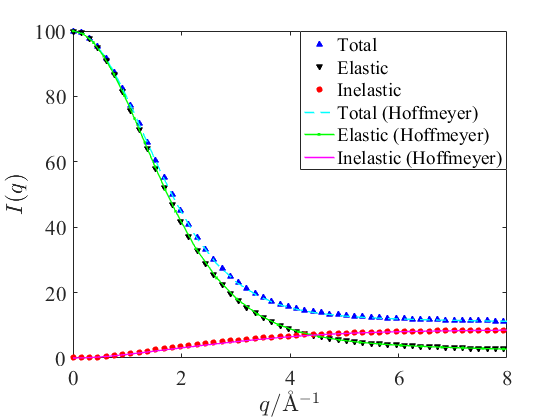}
        \caption{}
        \label{fig:Hoffmeyer1}
    \end{subfigure}
    \hfill
    \begin{subfigure}{1\columnwidth}
        \centering
        \includegraphics[width=1\textwidth]{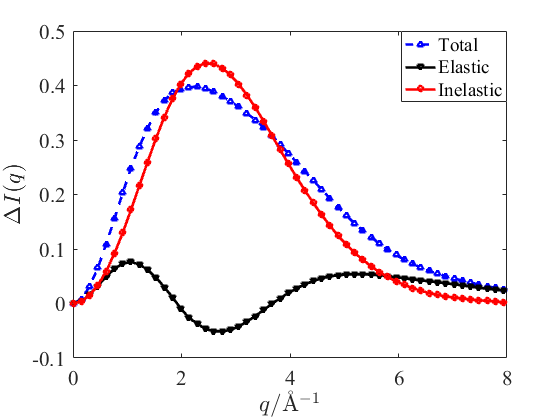}
        \caption{}
        \label{fig:Hoffmeyer2}
    \end{subfigure}\\
          \caption{Total, elastic and inelastic ground-state x-ray scattering curves of \ce{NH3}. The scattering curves are calculated with CASSCF(10,8)/aug-cc-pVTZ. The results are compared with previous calculations by Hoffmeyer \etal\ using MR-SDCI wavefunction ([5s3p2d/3s2p]) numerically integrated on a grid.~\cite{Hoffmeyer1998} Subfigure~\ref{fig:Hoffmeyer1} shows the total intensity, while subfigure~\ref{fig:Hoffmeyer2} shows the difference.}
    \label{fig:Hoffmeyer}
\end{figure*}

In Fig.~\ref{fig:Hoffmeyer} we show total, elastic and inelastic scattering curves of \ce{NH_3} at the reference CASSCF(10,8)/aug-cc-pVDZ level. The comparison with previous MR-SDCI (Multi Reference Single and Doubles Configuration Interaction) calculations by Hoffmeyer \etal~\cite{Hoffmeyer1998} shows rather good agreement given the differences in the methods, levels of theory and basis sets. Their approach relies on numerical integration, whereas our result is strictly analytical. Our best calculation employs aug-cc-pVDZ, while their work reports a smaller double-zeta basis set with polarization and diffuse functions, [5s3p2d/3s2p]. Unlike MR-SDCI, CASSCF calculations account only for static electron correlation. The dynamic electron correlation has a smaller effect on the elastic component of the scattering signal as seen by comparing to the MR-SDCI results. Its influence increases when total and inelastic scattering are considered, demonstrating the importance of electron correlation for these quantities. A systematic study of the effect of electron correlation is critically important for fully understanding gas-phase X-ray scattering experiments and will be addressed in subsequent work.

%
%


\subsection{Excited state total scattering}

 
 \begin{table}[htbp]
\caption{The ground state equilibrium geometry and geometries along the umbrella normal mode in the first excited state. The geometries are presented in terms of the \ce{N-H} bond distance and the pyramidization angle defined as the angle between the plane of the H atoms and any of the \ce{N-H} bonds. The equilibrium geometries in both states are given bold font.}
\begin{tabular}{lcc}
\hline \hline 
State & \parbox[t]{3cm}{Pyramidization \\ angle}  & \ce{N-H} bond /~\AA \\ 
 \hline
Ground  & \textbf{23.5$^{\circ}$} & \textbf{1.022} \\ 
\hline 
\multirow{11}{*}{Excited} & \textbf{0.0$^{\circ}$} & \textbf{1.032} \\ 
 & 3.7$^{\circ}$ & 1.034 \\ 
 & 7.4$^{\circ}$ & 1.041 \\ 
 & 11.0$^{\circ}$  & 1.051 \\ 
 & 14.6$^{\circ}$  & 1.066 \\ 
 & 18.0$^{\circ}$  & 1.085 \\ 
 & 21.3$^{\circ}$  & 1.108 \\ 
 & 24.5$^{\circ}$  & 1.133 \\ 
 & 27.5$^{\circ}$  & 1.163 \\ 
 & 30.3$^{\circ}$  & 1.196 \\ 
 & 33.0$^{\circ}$  & 1.231 \\ 
 \hline 
\end{tabular}
\label{table:geom}
\end{table}

Here, we consider a simple model that illustrates both the utility of our methodology and the nature of the signal detected in ultrafast x-ray scattering. For our purpose, a suitable candidate is the photoexcitation of ammonia to the first excited singlet state, whose initial dynamics follows an umbrella motion~\cite{Walsh1961}. The goal is to track the changes and dominant contributions to the total scattering as the geometry changes. In order to achieve that, we first optimize the ground state ion geometry at the CASSCF(9,8)/6-31+G* level as an approximation for the first excited state, which has Rydberg $3s$ character. The normal modes are calculated. The molecular geometry is then displaced in a series of steps along the umbrella mode and at each geometry the ground and the first excited states are calculated in a state-average fashion at SA2-CASSCF(10,8)/aug-cc-pVTZ level of theory. Equilibrium geometries and geometrical parameters along the displacement are presented in Table~\ref{table:geom}. The total and elastic scattering signals are computed for each state. The signals are then expressed as a fractional intensity change as it is commonly done in experiments (see e.g.\ Refs.\ \citenum{Minitti2015,BudarzJPB2015,Yong2018,Ruddock2019}),
\begin{equation} \label{eq:deltaS_2}
     \Delta S^{\mathrm{tot}}(q,\bar{\boldsymbol{R}}) = \frac{I^{\mathrm{tot}}_{\mathrm{exc}}(q,\bar{\boldsymbol{R}})-I^{\mathrm{tot}}_{\mathrm{gs}}(q,\bar{\boldsymbol{R}}_{0})}{I^{\mathrm{tot}}_{\mathrm{gs}}(q,\bar{\boldsymbol{R}}_{0})},
\end{equation}
where $I^{\mathrm{tot}}_{\mathrm{exc}}(q,\bar{\boldsymbol{R}})$ and $I^{\mathrm{tot}}_{\mathrm{gs}}(q,\bar{\boldsymbol{R}}_{0})$ are the total scattering intensities for the excited state and the ground state, respectively. This expression gives the change of the signal for the excited state at a specific geometry, $\bar{\boldsymbol{R}}$, relative to the scattering from the ground state at its equilibrium geometry, $\bar{\boldsymbol{R}}_0$, under the assumption that there is no geometry change in the ground state upon excitation. In order to investigate its underlying contributions, it is conceptually useful to rewrite the expression in Eq.\ (\ref{eq:deltaS_2}) as sum of two contributions,
\begin{align} \label{eq:deltaS_split1}
    \begin{split}
    &
     \Delta S^{\mathrm{tot}}(q,\bar{\boldsymbol{R}}) 
    =
     \Delta S^{\mathrm{tot}}_{\mathrm{elec}}(q,\bar{\boldsymbol{R}}) + \Delta S^{\mathrm{tot}}_{\mathrm{nucl}}(q,\bar{\boldsymbol{R}})
    \\
    & = 
    \frac{I^{\mathrm{tot}}_{\mathrm{exc}}(q,\bar{\boldsymbol{R}})-I^{\mathrm{tot}}_{\mathrm{gs}}(q,\bar{\boldsymbol{R}})}{I^{\mathrm{tot}}_{\mathrm{gs}}(q,\bar{\boldsymbol{R}}_{0})}
    +
    \frac{I^{\mathrm{tot}}_{\mathrm{gs}}(q,\bar{\boldsymbol{R}})-I^{\mathrm{tot}}_{\mathrm{gs}}(q,\bar{\boldsymbol{R}}_{0})}{I^{\mathrm{tot}}_{\mathrm{gs}}(q,\bar{\boldsymbol{R}}_{0})}.
     \end{split}
 \end{align}
The first term, henceforth called electronic, shows the difference solely due to the electronic redistribution at any given geometry. The second nuclear term indicates the contribution due to structural changes, and is defined with respect to the electronic structure of the ground state only. Furthermore, making use of the fact that the total scattering is a sum of elastic and inelastic scattering, each of these terms can be split into two contributions, 
\begin{align} \label{eq:deltaS_split2}
    \begin{split}
        &  \Delta S^{\mathrm{tot}}_{\mathrm{elec}}(q,\bar{\boldsymbol{R}}) = 
         \Delta S^{\mathrm{e}}_{\mathrm{elec}}(q,\bar{\boldsymbol{R}}) +  \Delta S^{\mathrm{i}}_{\mathrm{elec}}(q,\bar{\boldsymbol{R}}) 
        \\
        &  \Delta S^{\mathrm{tot}}_{\mathrm{nucl}}(q,\bar{\boldsymbol{R}}) = 
         \Delta S^{\mathrm{e}}_{\mathrm{nucl}}(q,\bar{\boldsymbol{R}}) +  \Delta S^{\mathrm{i}}_{\mathrm{nucl}}(q,\bar{\boldsymbol{R}}) .
     \end{split}
 \end{align}

\begin{figure*}[htb]
    \begin{subfigure}{1\columnwidth}
        \centering
        \includegraphics[width=1\textwidth]{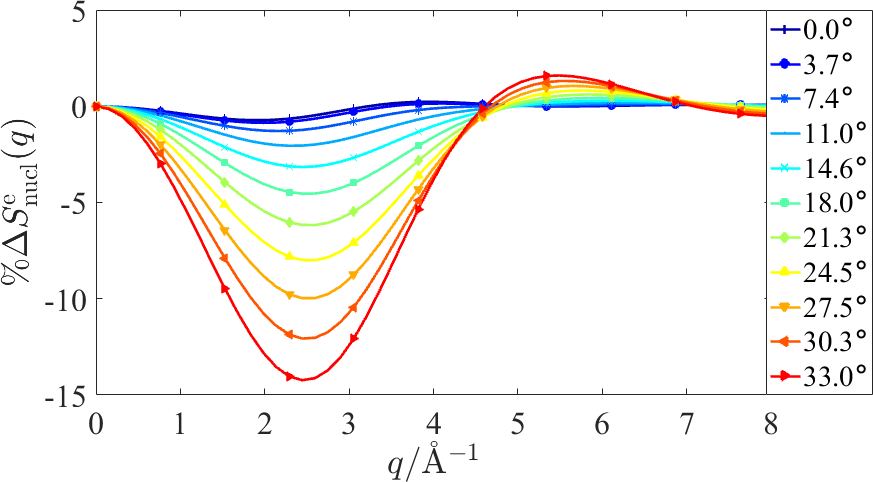}
        \caption{Nuclear elastic, $\% \Delta S^{\mathrm{e}}_{\mathrm{nucl}}(q)$ }
        \label{fig:A}
    \end{subfigure}
    \hfill
    \begin{subfigure}{1\columnwidth}
        \centering
        \includegraphics[width=1\textwidth]{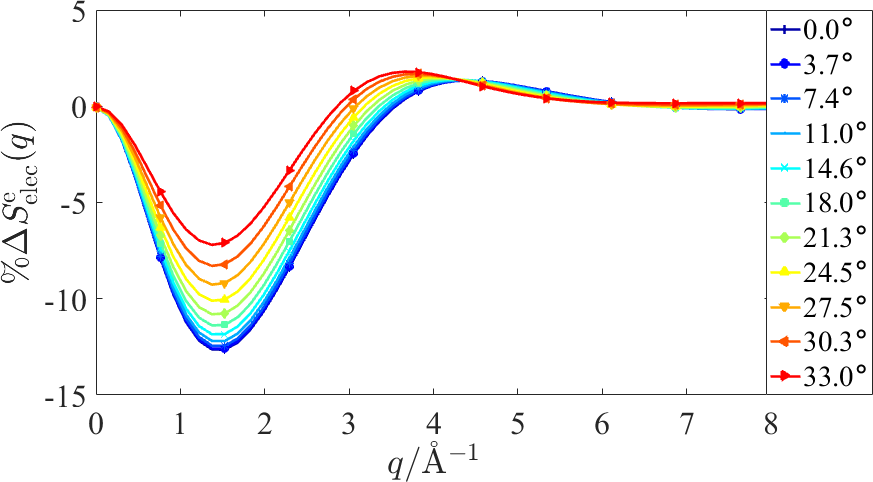}
        \caption{Electronic elastic, $\% \Delta S^{\mathrm{e}}_{\mathrm{elec}}(q)$}
        \label{fig:B}
    \end{subfigure}\\
    \begin{subfigure}{1\columnwidth}
        \centering
        \includegraphics[width=1\textwidth]{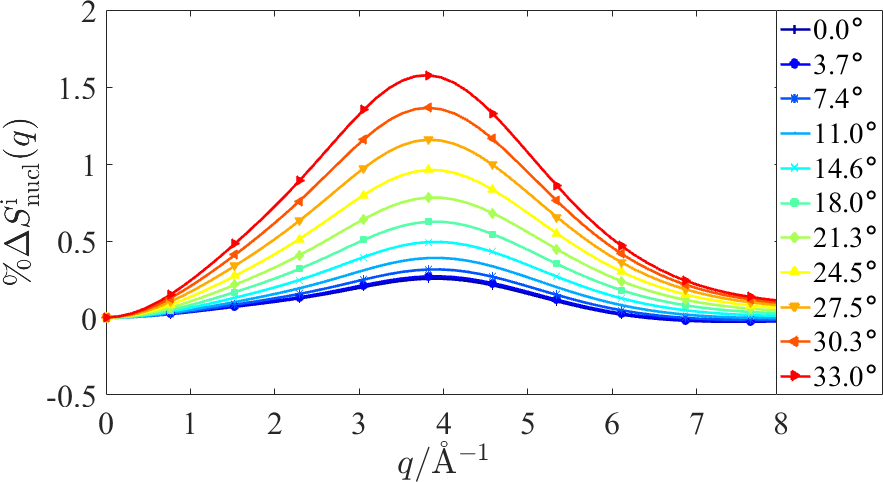}
        \caption{Nuclear inelastic, $\% \Delta S^{\mathrm{i}}_{\mathrm{nucl}}(q)$}
        \label{fig:C}
    \end{subfigure}
    \hfill
        \begin{subfigure}{1\columnwidth}
        \centering
        \includegraphics[width=1\textwidth]{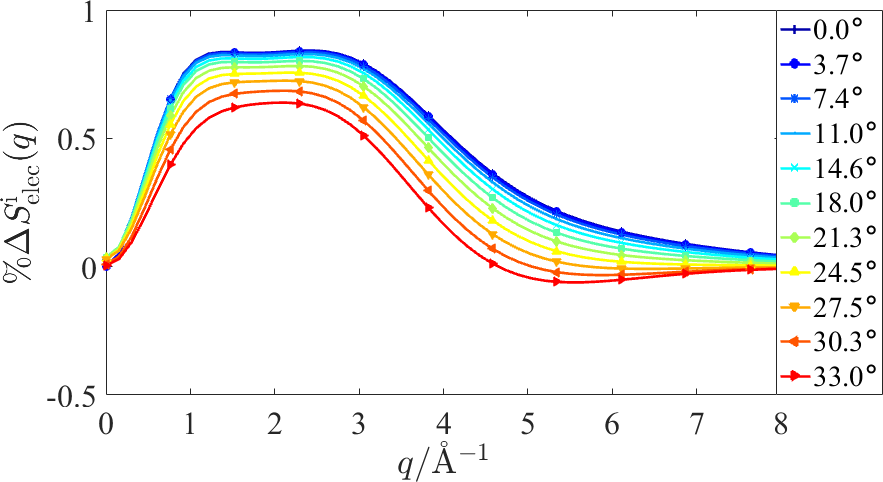}
        \caption{Electronic inelastic, $\% \Delta S^{\mathrm{i}}_{\mathrm{elec}}(q)$}
        \label{fig:D}
    \end{subfigure} \\
    \begin{subfigure}{1\columnwidth}
        \centering
        \includegraphics[width=1\textwidth]{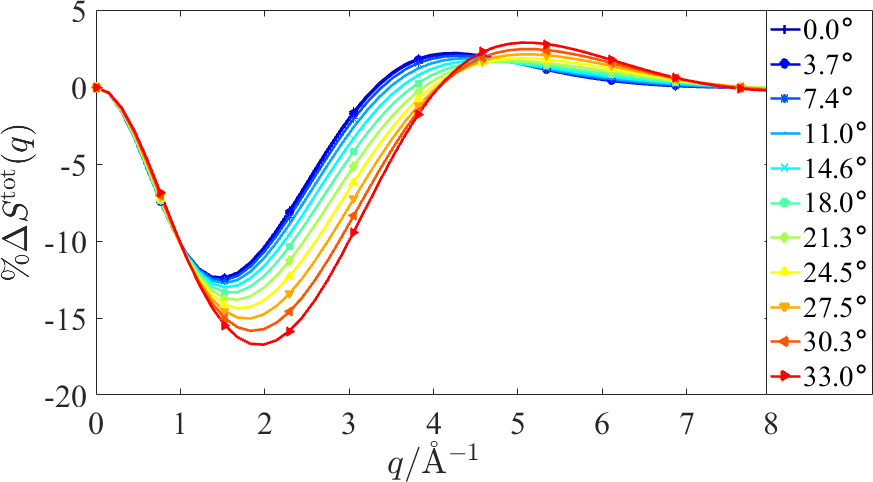}
        \caption{Total scattering, $\% \Delta S^{\mathrm{tot}}(q)$}
        \label{fig:E}
    \end{subfigure}
\caption{Breakdown of the total percentage fractional intensity change along the umbrella mode in the 3s Rydberg state of ammonia relative to the ground state equilibrium geometry. The displacement along the normal mode is labeled in terms of the pyramidization angle between an \ce{N-H} bond and the plane of the three hydrogen atoms. The nuclear elastic (a) and nuclear inelastic (c) terms indicate the changes due to the geometry evolution in the elastic and inelastic scattering, respectively. Similarly, the electronic elastic (b) and inelastic (d) components show the changes in scattering as a result of the difference between the electronic structure of the ground and the excited state at each geometry. The total scattering, \textit{i.e.}\ the sum of (a)--(d), is given in (e).}
\label{fig:split}
\end{figure*}

The breakdown of the total signal in terms of these four components is given in Fig.~\ref{fig:split}. The umbrella motion is tracked from 0 to 33.0 degrees in the pyramidization angle formed between the plane of the hydrogen atoms and one of the \ce{N-H} bonds. The displacement along the normal mode is accompanied by a \ce{N-H} bond elongation from 1.03~\AA\ to 1.23~\AA, which is seen to be the dominant factor for the variability in the nuclear part of the signal. The ground state equilibrium geometry used for this work has a pyramidization angle of 23.5 degrees and an \ce{N-H} bond length of 1.02~\AA. This implies that the planar geometry of the excited state is more similar to the ground state geometry, with referene to the \ce{N-H} distance that dominates the scattering. This is reflected by the small magnitude of the nuclear scattering in Fig.~\ref{fig:A} for this geometry. As the pyramidization angle increases, so does the bond length, which ultimately results in a maximum amplitude of $\% \Delta S$ of about 14\%. 

Meanwhile, the elastic part of the electronic contribution in Fig.~\ref{fig:B} shows much less variability with the change of the geometry. The overall shape seems to change very little and the maximum amplitude of the central peak ranges only from 7\% to 13\%. It is of paramount importance to point out that the magnitude of the electronic component is comparable to that attributed to the nuclear motion. While at large nuclear pyramidization angles, the nuclear contribution is approximately two times larger, at near-planar geometries the elastic signal is almost exclusively attributed to the redistribution of the electronic density in the Rydberg state. Given the range of momentum transfer vectors for which the dip in Fig.~\ref{fig:B} occurs, the observation can be explained by the effective loss of electron density in the molecular core associated with the delocalization of the Rydberg electron in the excited state. The small changes in the electronic component along the umbrella mode align with the fact that Rydberg electrons are not strongly affected by structural evolution of the ion-like core.

As expected, inspection of Figs.~\ref{fig:C} and \ref{fig:D} reveals that the magnitude of the inelastic scattering is generally about 10 times smaller than the corresponding elastic contribution. Nonetheless, the size of the inelastic contribution is clearly large enough to have a tangible effect on interpretation of the experimental data. Here, the patterns parallel those seen in the elastic scattering. The electronic contribution is relatively constant with a magnitude comparable to the nuclear effect. The largest pyramidization angle has the largest nuclear component and the smallest electronic, with about twofold difference between the two. The planar \ce{NH3} is dominated by the electronic scattering. Interestingly, the geometry dependence of the inelastic scattering is rarely accounted for in experiments. Specifically, the inelastic scattering is often approximated as an incoherent sum of stationary inelastic Compton factors for individual atoms. This is clearly a poor approximation in this case and most likely in general when considering small effects in the fractional signal change in time-resolved gas-phase experiments.~\cite{Carrascosa2019total} 
Given that the real experimental observable is the total scattering that stems from the Fourier transform of the two-electron electron density, it seems natural to consider the inelastic effect on an equal footing with elastic scattering. As seen here, inelastic scattering can account for up to 10\% of the fractional signal change partially attributed to the difference between the two states considered, and exhibits a geometry-dependence similar to elastic scattering.

When pumped into the excited state, ammonia undergoes fast umbrella motion. The picture that emerges from this work is that the observed signal will alternate between two extrema driven by the elongation and contraction of the \ce{N-H} distance. However, as shown in Fig.~\ref{fig:E}, the baseline for this oscillation is set by the shape of the relatively constant difference in the scattering signals between the Rydberg state and the ground state. The latter is attributed to the electron-density hole brought about by the promotion of an electron from the HOMO to the diffuse 3s Rydberg orbital. The inelastic scattering has smaller but far from negligible effect on the scattering signal.

\section{Conclusions}
The mathematical framework and computational approach presented in this article allows for the efficient calculation of the isotropic total, elastic, inelastic and coherent mixed scattering signals. In order to perform the integration over the Euler angles needed to achieve spherical averaging, we consider \abin wavefunctions expressed in a basis of Gaussian-type orbitals. Analytic solutions to the Fourier transform from real to reciprocal space results in a series of products of Gaussian and spherical Bessel functions, which are relatively easy to evaluate computationally. The approach is benchmarked against previous numerical calculations in the case of ammonia. We demonstrate the scaling of the algorithms with the basis sets and levels of theory used.

The methodology described in the paper is utilized to investigate a simplified model of the photoexcitation of ammonia to its $3s$ Rydberg state. The observed elastic scattering signal shows a strong signature of the shift in the electron density associated with the promotion of an electron from the HOMO to the $3s$ Rydberg orbital. The magnitude of this purely electronic effect is comparable to the geometry dependent part of the signal. In addition, ammonia shows strong change in the inelastic scattering upon excitation, which is driven by changes in the electronic structure. It was furthermore shown that changes in geometry also play a role in inelastic scattering.

As ultrafast gas-phase x-ray scattering experiments are becoming more and more successful in obtaining high-quality data, it is of paramount importance to have the right tools to analyze the results. The internal dynamics, both nuclear and electronic, is encoded in the isotropic part of the signal, which can be extracted by means of Legendre decomposition of the detector signal. The isotropic signal should be understood as the spherical average of the Fourier transform of the correlated two-electron density of the molecule. The approach presented here allows this signal to be calculated efficiently for ground and excited states and can be used to aid the interpretation of pump-probe ultrafast x-ray scattering experiments. 

Going further, a similar mathematical apparatus can be applied to the case of static molecules or arbitrary high-order terms in the Legendre decomposition of the fully dimensional signal. The former can be achieved readily by forgoing the Spherical Bessel function expansion, while the latter can be achieved, albeit in a less straightforward manner, by using higher-order spherical Bessel functions. Equally important is the question of the \abin level of theory used to calculate the molecular wavefunctions. Given that the signal is related to the two-electron density, it can be expected that the total scattering shows high sensitivity to electron correlation. Hence, thorough investigation of the impact of the post-Hartee-Fock methods is urgently needed.

\begin{acknowledgement}
N.Z.\ acknowledges a Carnegie Ph.D.\ Scholarship. A.K.\ acknowledges support from a Royal Society of Edinburgh Sabbatical Fellowship (58507) and a research grant from the Carnegie Trust for the Universities of Scotland (CRG050414). 
\end{acknowledgement}

\bibliography{references}

\end{document}